\documentclass[twocolumn,trackchanges]{aastex63}
\usepackage{amsmath}
\usepackage{CJK}
\usepackage{color}
\received{May 8, 2020}
\revised{August 31, 2020}
\accepted{September 9, 2020}
\submitjournal{Astronomical Journal}

\shorttitle{Small Planet Occurrence Increases with Metallicity for
Late-type Dwarfs}
\shortauthors{Lu et al. (2020)}

\begin{document}
\begin{CJK*}{UTF8}{gbsn}

\title{An Increase in Small-planet Occurrence with Metallicity for
Late-type Dwarf Stars in the Kepler Field and Its Implications for
Planet Formation}

\correspondingauthor{Cicero X. Lu}
\email{cicerolu@jhu.edu}

\author[0000-0001-9352-0248]{Cicero X. Lu}
\affiliation{Department of Physics and Astronomy, Johns Hopkins
University, 3400 N Charles St, Baltimore, MD 21218, USA}

\author[0000-0001-5761-6779]{Kevin C. Schlaufman}
\affiliation{Department of Physics and Astronomy, Johns Hopkins
University, 3400 N Charles St, Baltimore, MD 21218, USA}

\author[0000-0002-9156-7461]{Sihao Cheng (程思浩)}
\affiliation{Department of Physics and Astronomy, Johns Hopkins
University, 3400 N Charles St, Baltimore, MD 21218, USA}

\begin{abstract}

\noindent
While it is well established that giant-planet occurrence rises rapidly
with host star metallicity, it is not yet clear if small-planet occurrence
around late-type dwarf stars depends on host star metallicity.  Using the
Kepler Data Release 25 planet candidate list and its completeness data
products, we explore planet occurrence as a function of metallicity
in the Kepler field's late-type dwarf stellar population.  We find
that planet occurrence increases with metallicity for all planet radii
$R_{\mathrm{p}}$ down to at least $R_{\mathrm{p}}\approx2~R_{\oplus}$
and that in the range $2~R_{\oplus}\lesssim R_{\mathrm{p}}\lesssim
5~R_{\oplus}$ planet occurrence scales linearly with metallicity $Z$.
Extrapolating our results, we predict that short-period planets with
$R_{\mathrm{p}}\lesssim 2~R_{\oplus}$ should be rare around early M
dwarf stars with $\mathrm{[M/H]}\lesssim-0.5$ or late M dwarf stars with
$\mathrm{[M/H]}\lesssim+0.0$.  This dependence of planet occurrence on
metallicity observed in the Kepler field emphasizes the need to control
for metallicity in estimates of planet occurrence for late-type dwarf
stars like those targeted by Kepler's K2 extension and the Transiting
Exoplanet Survey Satellite (TESS).  We confirm the theoretical expectation
that the small planet occurrence--host star metallicity relation is
stronger for low-mass stars than for solar-type stars.  We establish
that the expected solid mass in planets around late-type dwarfs in the
Kepler field is comparable to the total amount of planet-making solids
in their protoplanetary disks.  We argue that this high efficiency of
planet formation favors planetesimal accretion over pebble accretion
as the origin of the small planets observed by Kepler around late-type
dwarf stars.

\end{abstract}

\keywords{Exoplanet formation (492); Exoplanets (498); Extrasolar rocky
planets (511); Extrasolar ice giants (2024); Late-type dwarf stars (906);
Planet hosting stars (1242)}

\section{Introduction}

Planet formation is seeded by the metals present in a protoplanetary
disk.  It must be the case that the total heavy-element content of
a protoplanetary disk provides an upper limit on the solid mass
of planets formed in that disk.  For that reason, there must be
a metallicity below which even Earth-mass planets cannot form.
Observations have shown that the occurrence of giant planets
around FGKM dwarf stars rises rapidly with host star metallicity
\citep[e.g.,][]{Santos2004,Fischer+Valenti2005,Johnson+Apps2009,Johnson+2010}.
However, it is not clear how host star metallicity influences small
planet occurrence.

Since both the metallicity and mass of a protoplanetary disk determine
the amount of planet-making material available, we expect that the
dependence of small-planet occurrence on metallicity should be stronger
for low-mass stars than for solar-type stars.  A star has accreted
the vast majority of the total mass initially in its young disk prior
to the epoch of planet formation, so any residual solids locked up in
planets will not affect the observed metallicity of a star.  Since the
star and disk both formed from the same molecular core, the overall
metallicities $Z_{\star}$ and $Z_{\mathrm{disk}}$ should be the same.
The mass of a disk during the epoch of planet formation has been found
to scale roughly linearly with stellar mass with fixed disk-to-star
mass ratio $M_{\mathrm{disk}}/M_{\star}$ in the range $0.002 \lesssim
M_{\mathrm{disk}}/M_{\star} \lesssim 0.006$ \citep[e.g.,][]{Andrews2013}.
Though there is more than an order-of-magnitude scatter in the
\citet{Andrews2013} relation, those authors favor an inherently linear
relationship between $M_{\mathrm{disk}}$ and $M_{\star}$.  The net
effect is that the amount of planet-making material available in a
protoplanetary disk should scale roughly linearly with both stellar
metallicity $Z_{\star}$ and mass $M_{\star}$.

Assuming the solar metallicity $Z_{\odot} = 0.014$
\citep[e.g.,][]{Asplund+09} and a disk-to-star mass ratio
$M_{\mathrm{disk}}/M_{\star} = 0.01$ \citep{Andrews2013}, an early M dwarf
with $M_{\star} = 0.6~M_{\odot}$ and $[\mathrm{M/H}] = -0.5$ would have
less than $10~M_{\oplus}$ of planet-making material in its disk.  In that
case, the available planet-making material is much less than the amount
required to make a single Neptune-size planet with radius $R_{\mathrm{p}}
= 4~R_{\oplus}$ \citep[Neptune has at least $13~M_{\oplus}$ of metals
as shown by][]{Podolak+19}.  On the other hand, $10~M_{\oplus}$ of
planet-making material would be enough to make an Earth-composition
super-Earth mass planet with $R_{\mathrm{p}} \lesssim 1.8~R_{\oplus}$
\citep[e.g.,][]{Zeng+2019}.  If the timescale for growing Earth-mass
embryos scales with the amount of planet-making material as suggested
by detailed calculations \citep{Movshovitz+10}, then the probability
of forming a planet with a significant gaseous envelope in the few Myr
available before its parent protoplanetary disk is dissipated should
also scale with the amount of planet-making material.  There are hints
that this effect becomes important at $[\mathrm{M/H}] \approx -0.5$
for solar-type stars \citep[e.g.,][]{Petigura+18}.

The expected relationship between small-planet occurrence
and metallicity for low-mass stars has been hard to confirm
because metallicity measurements for low-mass stars are
inherently difficult.  Stellar metallicity has traditionally
been measured using metal lines in optical or near-infrared spectra
\citep[e.g.,][]{Rojas-Ayala+2010,Rojas-Ayala+2012,Mann+2013Metal,Mann+2013KeplerKM_Non,Muirhead+2014,Neves+2014,Newton+2014}.
Metal lines and molecular absorption bands are so common in the optical
spectra of cool dwarf stars that it often becomes impossible to set
the continuum level necessary for the measurement of equivalent widths.
The lack of laboratory data necessary to handle molecular features has
been an issue as well.

Broadband photometry also carries metallicity information, albeit at a
less precise level for individual stars.  Applied to large samples of
stars in the same place on the sky distributed over a similar range
in distance, photometric metallicities become precise indicators of
relative metallicity.  \citet{Bonfils+2005} and \citet{Johnson+2009}
were among the first to compare the photometric metallicities of
late-type dwarf stars observed to host or lack planets discovered with
the Doppler technique with the goal to explore the connection between
planet occurrence and metallicity.  \citet{Schlaufman+Laughlin2010}
built on these groundbreaking studies and found a hint that M dwarfs
hosting Neptune-mass planets are more metal-rich than similar stars
without planets.  Leveraging the large number of small planets discovered
early in the Kepler mission, \citet{Schlaufman+Laughlin2011} found that
the average $g-r$ of late-type dwarf stars with small-planet candidates
was $4\sigma$ redder than the average color of a control sample of
similar stars without identified planet candidates.  They argued that
their observation was evidence for a metallicity difference between
late-type dwarf stars with and without small planets.

The \citet{Schlaufman+Laughlin2011} result was criticized by
\citet{Mann+2012,Mann+2013KeplerKM_Non}, who argued that the $g-r$
photometric metallicity indicator used by \citet{Schlaufman+Laughlin2011}
is insensitive to metallicity and that the possible presence of giant
stars mistaken for dwarf stars in the \citet{Schlaufman+Laughlin2011}
control sample could produce a similar $g-r$ offset unrelated to
metallicity.  Both of these criticisms can now be conclusively
addressed.  Photometric metallicity relations for late-type
dwarf stars calibrated by reliable APO Galactic Evolution
Experiment (APOGEE) high-resolution $H$-band spectroscopy are now
available \citep{Majewski16,Schmidt2016}.  Kepler asteroseismology
\citep{Hekker+2011,Huber+2011,Stello+2013,Huber+2014,Mathur+2016,Yu+2016,Yu+18}
and Gaia DR2 parallaxes
\citep{Gaia16,Gaia18,Arenou+2018,Hambly+2018,Lindegren+2018,Luri+2018}
enable the construction of samples of dwarf stars without planet
candidates completely free of subgiant or giant star contamination.
Advances in the analysis of Kepler data and the public availability of
its completeness data products now permit differential planet occurrence
calculations.  Thanks to these developments, photometric metallicities
have become a powerful tool for the exploration of the small-planet
occurrence--metallicity relation.

Advances in the theory of planet formation have also revealed the
possible significance of the accretion of ``pebbles'', or material
significantly smaller than the km-size planetesimals historically studied
\citep{Ormel+10,Lambrechts+12}.  This ``pebble accretion'' process invokes
the accretion by planetary embryos of small particles experiencing strong
aerodynamic drag.  Since not all of this rapidly migrating material
can be accreted by a planetary embryo, pebble accretion is inherently
lossy in the sense that more than 90\% of a disk's initial complement of
planet-making material falls onto its host star \citep[e.g.,][]{Lin+18}.
On the other hand, the classical ``planetesimal accretion'' process is
thought to be much more efficient in the sense that a larger fraction
of a disk's initial complement of planet-making material is locked up in
planetesimals.  Efficient planet formation seems to have occurred in the
solar system, as the amount of planet-making material in the minimum-mass
solar nebula \citep[MMSN -][]{Weidenschilling1977,Hayashi1981} is within
a factor of two of that expected in a disk with $M_{\mathrm{disk}}
\sim 0.01~M_{\odot}$ and $Z_{\mathrm{disk}} \sim Z_{\odot} = 0.014$.
We therefore propose that the efficiency of planet formation---the
fraction of a protoplanetary disk's initial complement of planet-making
material sequestered in planets---is diagnostic of the relative importance
of pebble/planetesimal accretion in the planet formation process.

As we will show, it is now possible to use photometric metallicities to
explore small-planet occurrence around late-type dwarf stars as a function
of metallicity using Kepler data.  In this paper, we calculate planet
occurrence as a function of metallicity, orbital period, and planet
radius in the population of late-type dwarf stars observed by Kepler
during its prime mission.  Using the Kepler Data Release (DR) 25 Kepler
Object of Interest (KOI) planet candidate list \citep{Thompson2018},
we find that that planet occurrence increases with metallicity for all
planet radii down to at least $R_{\mathrm{p}} \approx 2~R_{\oplus}$
and that in the range $2~R_{\oplus} \lesssim R_{\mathrm{p}} \lesssim
5~R_{\oplus}$ planet occurrence scales linearly with metallicity.
In Section~\ref{section:data} we discuss our sample selection, describe
the photometric effective temperature and metallicity relations we use,
and outline the process we use to remove giant stars from our sample of
stars without planet candidates.  We split both planet candidate-host
and non-planet-candidate-host samples into metal-rich and metal-poor
subsamples and illustrate two different occurrence calculations in
Section~\ref{section:analysis}.  We then calculate planet formation
efficiency in an attempt to infer the relative importance of planetesimal
accretion and pebble accretion.  In Section \ref{section:discussion}
we discuss our results and their implications for the theory of
planet formation.  We conclude and summarize our findings in Section
\ref{section:conclusion}.

\section{Data}\label{section:data}

We seek to assemble the sample of late-type dwarf stars with Kepler
light curves that have been searched for transiting planet candidates.
To do so, we select late-type stars from the Kepler Input Catalog
\citep[KIC -][]{Brown2011} with effective temperature $T_{\mathrm{eff}}$
in the range 3600 K $\lesssim T_{\mathrm{eff}} \lesssim$ 4200 K using
the following empirical relations from \citet{Schmidt2016} based on
spectroscopic stellar parameters derived from APOGEE high-resolution
$H$-band spectroscopy
\begin{eqnarray}
[\mathrm{M/H}] & = & a_0 + a_1 \left(r-z\right) + a_2 \left(W1 -W2\right),\label{mh_relation}\\
T_{\mathrm{eff}} & = & b_0 + b_1 \left(r-z\right) + b_2 \mathrm{[M/H]},\label{teff_relation}
\end{eqnarray}
with the coefficients $a_i = (-0.822, 0.634, -4.508)$ and $b_i = (4603.4,
-576.5, 225.0)$.  \citet{Schmidt2016} considered all color combinations
possible with Sloan Digital Sky Survey (SDSS) $ugriz$, Two Micron All Sky
Survey \citep[2MASS -][]{Skrutskie+06} $JHK_{\mathrm{s}}$, and Wide-field
Infrared Survey Explorer \citep[WISE -][]{Wright+10,Mainzer+11} $W1$
and $W2$ photometry.  They found that a linear relation using $r-z$ and
$W1-W2$ was best able to reproduce the spectroscopic stellar parameters
$T_{\mathrm{eff}}$ and $\mathrm{[M/H]}$.  The uncertainties in individual
$\mathrm{[M/H]}$ and $T_{\mathrm{eff}}$ estimates produced using Equations
\eqref{mh_relation} and \eqref{teff_relation} are approximately 0.2 dex
in $[\mathrm{M/H}]$ and 100 K in $T_{\mathrm{eff}}$.  We require all
late-type stars in our sample to have been observed for at least one
quarter during the Kepler mission.

It is well known that the surface gravity $\log{g}$ estimates in the
KIC are imperfect \citep[e.g.,][]{Mann+2012,Dressing+2015}.  To ensure
that there are no giant stars in our sample, we reject stars identified
as giants via either asteroseismic oscillations or Gaia DR2 parallaxes.
We first select Kepler target stars with KIC $\log{g} > 4$.  We then
remove stars identified through asteroseismology as subgiants or as
giants/red clump stars by \citet{Hekker+2011}, \citet{Huber+2011},
\citet{Stello+2013}, \citet{Huber+2014}, \citet{Mathur+2016}, or
\citet{Yu+2016,Yu+18}.  We also use Gaia DR2 parallaxes to calculate
Gaia $G$-band absolute magnitudes and then exclude 11 giant stars that
are several magnitudes above the \citet{Hamer+19} empirical Pleiades
zero-age main sequence.

We cross match this purified sample of late-type dwarf stars with
the Kepler DR25 list of KOIs dispositioned as planet candidates
\citep{Thompson2018}.  We use the homogeneous DR25 planet candidate
list because it was generated in a fully automated fashion that
eliminated human vetting of threshold crossing events.  That lack of
intervention made its completeness straightforward to algorithmically
assess.  Because giant planet host stars are known to be metal rich, we
exclude from our analysis stars that host planets with $R_{\mathrm{p}}
> 5~R_{\oplus}$.  We also verified that using the updated stellar radii
from \citet{Berger+2018} did not change any of our subsequent conclusions.

\begin{figure*}
\plottwo{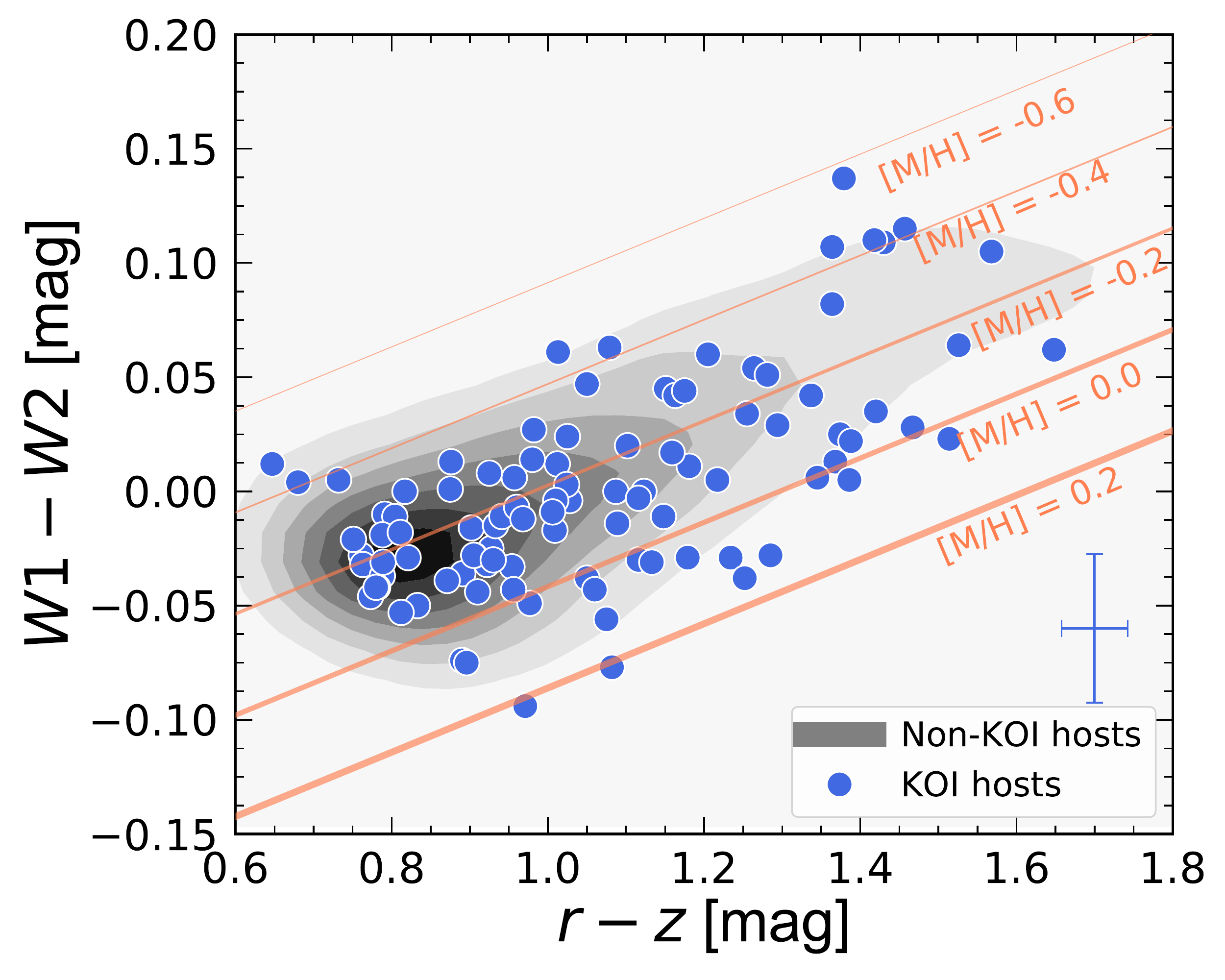}{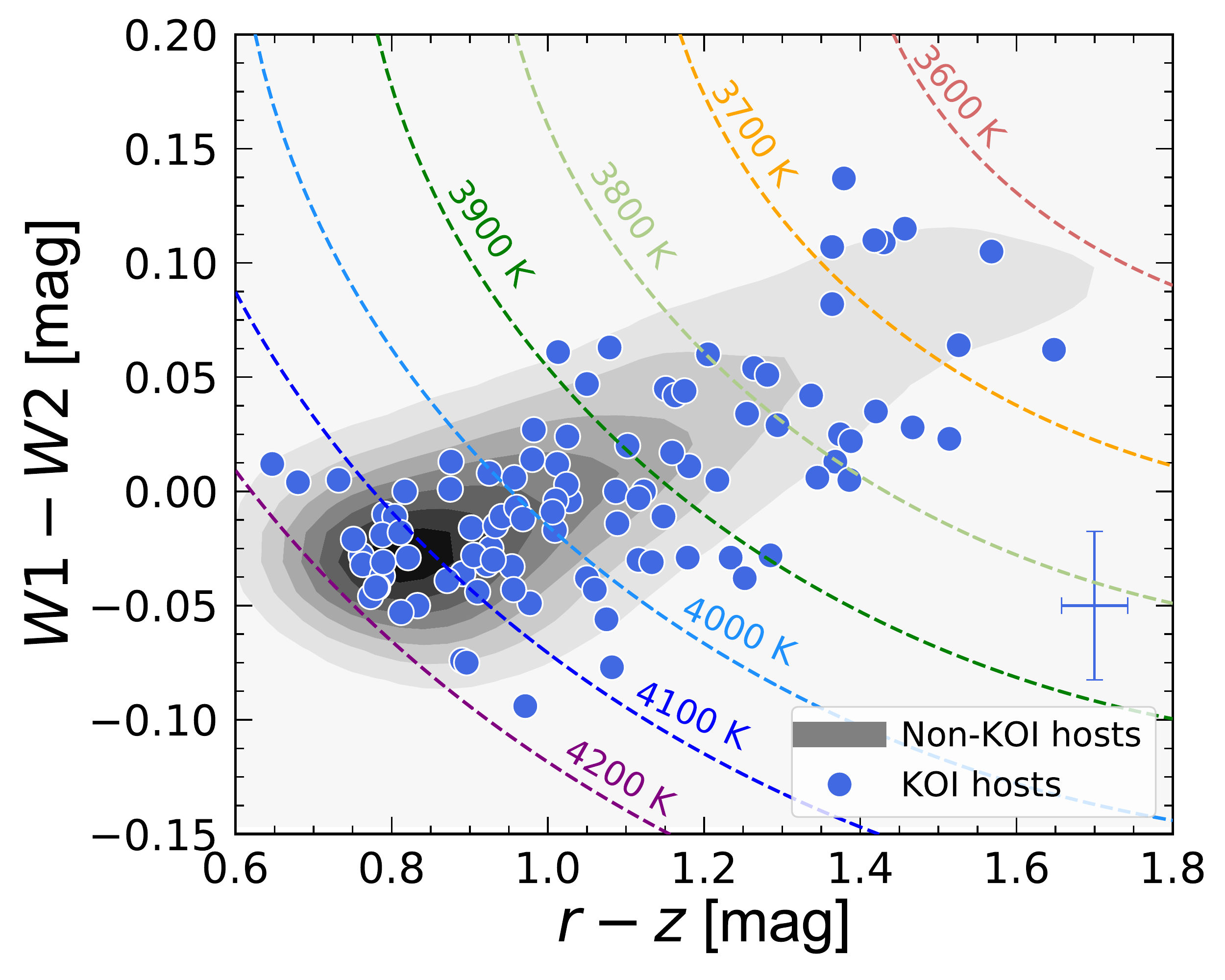}
\caption{Distribution of late-type dwarf stars in the $r-z$ versus $W1-W2$
color--color plot. The blue circles represent the planet candidate-host
sample while the gray shading represents the distribution of the
non-planet-candidate-host sample after kernel smoothing.  We represent
the typical color uncertainties with the crosses at the bottom right
of each plot.  Left: solid orange lines represent lines of constant
$[\mathrm{M/H}]$ according to \citet{Schmidt2016}.  Right: dashed
curves represent lines of constant $T_{\mathrm{eff}}$ according to
\citet{Schmidt2016}.\label{fig01}}
\end{figure*}

\begin{figure}
\plotone{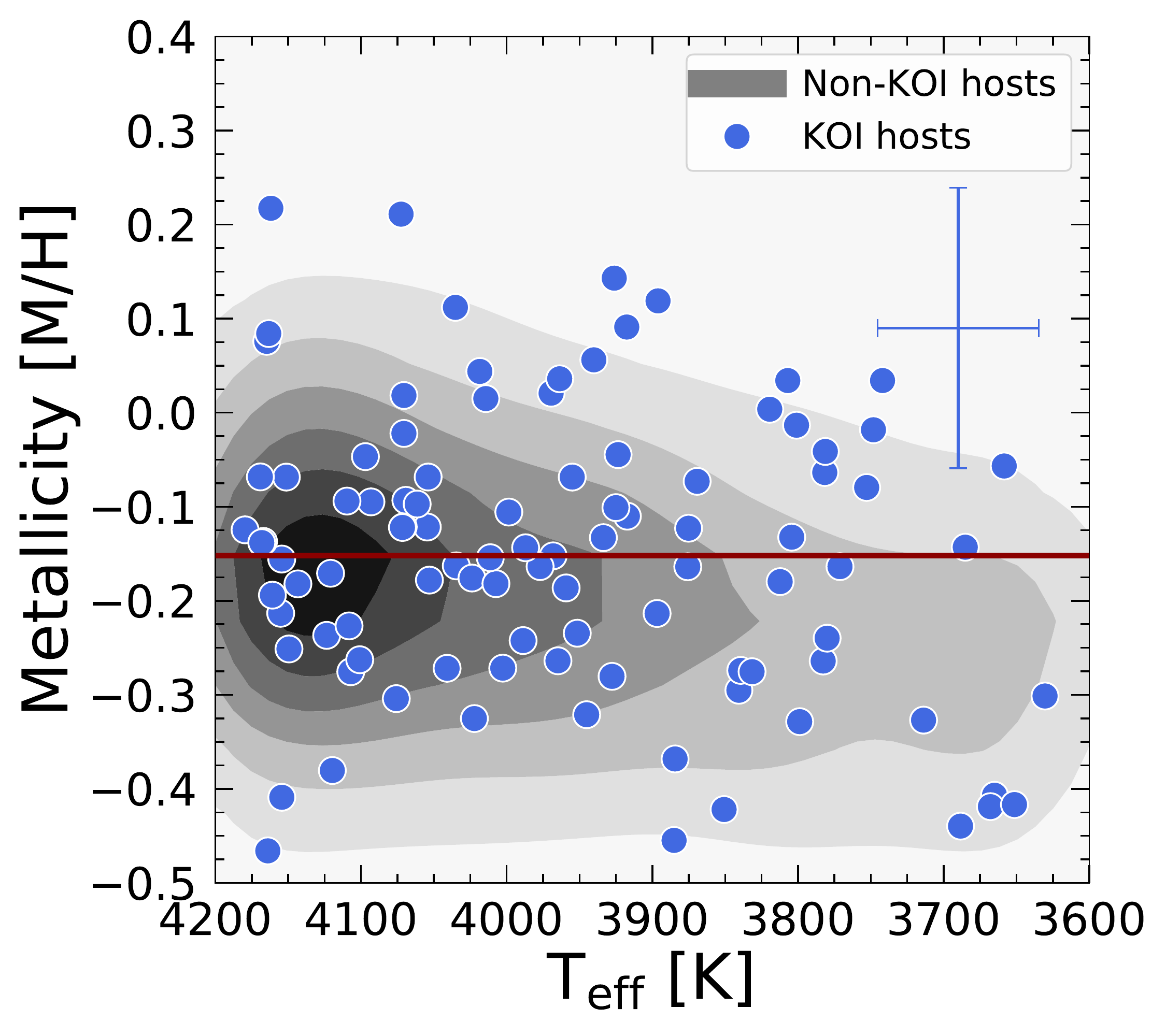}
\caption{$\mathrm{[M/H]}$ as a function of $T_{\mathrm{eff}}$ for planet
candidate-host and non-planet-candidate-host samples inferred using $r-z$
and $\mathrm{W1} - \mathrm{W2}$ colors in Equations \eqref{mh_relation}
and \eqref{teff_relation}.  The blue circles represent the planet
candidate-host sample while the gray shading represents the distribution
of the non-planet-candidate-host sample after kernel smoothing.
The red line at $[\mathrm{M/H}] = -0.15$ indicates the median host star
metallicity for the planet candidate sample.  While there are fewer stars
above the red line than below it, the metal-rich stars host more multiple
planet candidate systems.  We represent the typical $[\mathrm{M/H}]$
and $T_{\mathrm{eff}}$ uncertainties resulting from uncertainties in
the input photometry with the cross at the top right.\label{fig02}}
\end{figure}

Our final planet candidate-host sample consists of the 99 late-type
dwarfs with at least one planet candidate with $R_{\mathrm{p}} \leq
5~R_{\oplus}$ listed in Table~\ref{tbl-1}.  We refer to these stars as
our planet candidate-host sample from this point on.   We also select a
sample of 3,395 late-type dwarfs that were part of the main transiting
exoplanet search program, passed all of our selection criteria listed
above, and have no detected planet candidate.  We list these stars in
Table~\ref{tbl-2} and refer to them as our non-planet-candidate-host
sample from here.  We plot $r-z$ versus $W1-W2$ color--color plots for
both our planet candidate-host and non-planet-candidate-host samples
in Figure~\ref{fig01}.  We plot photometric $T_{\mathrm{eff}}$ and
$[\mathrm{M/H}]$ values inferred using Equations \eqref{mh_relation}
and \eqref{teff_relation} for both our planet candidate-host and
non-planet-candidate-host samples in Figure~\ref{fig02}.

\begin{deluxetable*}{ccccccccccC}
\tablecaption{Late-type Dwarf Kepler Targets with at Least One DR25
Planet Candidate with $R_{\mathrm{p}} \leq 5~R_{\oplus}$}
\tablehead{
\colhead{KIC Number} & \colhead{Kepler Name} & \colhead{KOI Name} &
\colhead{R.A.} & \colhead{Decl.} & \colhead{$r$} & \colhead{$z$} &
\colhead{$W1$} & \colhead{$\sigma_{W1}$} &
\colhead{$W2$} & \colhead{$\sigma_{W2}$}\\
\colhead{} & \colhead{} & \colhead{} & \colhead{(deg)} & \colhead{(deg)}
& \colhead{(mag)} & \colhead{(mag)} & \colhead{(mag)} & \colhead{(mag)}& \colhead{(mag)} & \colhead{(mag)}}
\startdata
10118816 &        \nodata &  K01085 &  281.05011 &  47.188148 &   15.29 &   14.28 &   12.164 &    0.023 &  12.103 &  0.021 \\
  6921944 &  Kepler-1105 &  K02114 &  281.11176 &  42.454910 &   15.14 &   14.31 &   12.164 &    0.023 &  12.193 &  0.022 \\
  7582691 &        \nodata &  K04419 &  281.11646 &  43.282440 &   15.16 &   14.29 &   12.145 &    0.023 &  12.132 &  0.022 \\
  6497146 &   Kepler-438 &  K03284 &  281.64581 &  41.951092 &   14.61 &   13.39 &   11.080 &    0.023 &  11.075 &  0.020 \\
  7870390 &    Kepler-83 &  K00898 &  282.23251 &  43.665630 &   15.76 &   14.96 &   12.880 &    0.023 &  12.891 &  0.024 \\
  8346392 &   Kepler-777 &  K01141 &  282.72406 &  44.346470 &   15.97 &   15.04 &   13.030 &    0.023 &  13.055 &  0.024 \\
  7094486 &  Kepler-1009 &  K01907 &  282.85992 &  42.665760 &   15.34 &   14.40 &   12.257 &    0.023 &  12.268 &  0.022 \\
 10386984 &   Kepler-658 &  K00739 &  282.98380 &  47.578590 &   15.52 &   14.63 &   12.535 &    0.023 &  12.571 &  0.022 \\
  7871954 &   Kepler-303 &  K01515 &  283.13547 &  43.657051 &   14.40 &   13.58 &   11.550 &    0.023 &  11.550 &  0.021 \\
 10122538 &  Kepler-1388 &  K02926 &  283.33606 &  47.174541 &   16.30 &   15.39 &   13.269 &    0.023 &  13.313 &  0.025 \\
\enddata
\tablecomments{The typical $r$- and $z$-band uncertainties are 0.02
mag \citep{Brown2011}.  This table is ordered by right ascension and is
available in its entirety in the machine-readable format.\label{tbl-1}}
\end{deluxetable*}

\begin{deluxetable*}{cccccccccc}
\tablecaption{Late-type Dwarf Kepler Targets with No Observed Planet Candidates}
\tablehead{
\colhead{KIC Number} & \colhead{R.A.} & \colhead{Decl.} & \colhead{$r$ } &
\colhead{$z$} & \colhead{$W1$} & \colhead{$\sigma_{W1}$}  & \colhead{$W2$} & \colhead{$\sigma_{W2}$}\\
\colhead{} & \colhead{(deg)} & \colhead{(deg)} & \colhead{(mag)} &
\colhead{(mag)} & \colhead{(mag)} & \colhead{(mag)} & \colhead{(mag)} & \colhead{(mag)}}
\startdata
7797376 &  279.708780 &  43.53535 &   15.78 &   15.14 &   13.108 &    0.024 &  13.106 &  0.025 \\
  7867105 &  279.996150 &  43.67716 &   14.32 &   13.53 &   11.488 &    0.023 &  11.520 &  0.021 \\
  7867279 &  280.120470 &  43.68729 &   15.39 &   14.70 &   12.614 &    0.023 &  12.642 &  0.023 \\
  7658133 &  280.122220 &  43.35148 &   15.52 &   14.80 &   12.790 &    0.022 &  12.796 &  0.023 \\
 10382584 &  280.132140 &  47.59633 &   15.85 &   14.93 &   12.756 &    0.023 &  12.793 &  0.024 \\
  7581219 &  280.137160 &  43.21006 &   15.97 &   14.68 &   12.352 &    0.023 &  12.331 &  0.021 \\
 10317398 &  280.232649 &  47.45096 &   15.03 &   14.39 &   12.458 &    0.023 &  12.436 &  0.022 \\
 10251684 &  280.305889 &  47.39492 &   15.09 &   14.36 &   12.340 &    0.022 &  12.354 &  0.022 \\
  7581487 &  280.315260 &  43.22374 &   15.76 &   14.62 &   12.377 &    0.023 &  12.357 &  0.022 \\
  7867585 &  280.358460 &  43.61532 &   16.01 &   15.01 &   12.824 &    0.023 &  12.821 &  0.023 \\
\enddata
\tablecomments{The typical $r$- and $z$-band uncertainties are 0.02
mag \citep{Brown2011}.  This table is ordered by right ascension and is
published in its entirety in the machine-readable format.\label{tbl-2}}
\end{deluxetable*}

\section{Analysis}\label{section:analysis}

We explore the connection between host star metallicity and small-planet
occurrence in three ways.  First, we use logistic regression to
estimate the significance of metallicity and effective temperature for
the prediction of planet occurrence in our complete sample.  We next
separate our complete sample into metal-rich and metal-poor subsamples
for which we independently calculate planet occurrence as a function of
metallicity, orbital period $P$, and planet radius $R_{\mathrm{p}}$ using
the Kepler DR25 completeness data products.  We then use a mass--radius
relation combined with the small-planet occurrence maps inferred for our
complete sample as well as our metal-rich and metal-poor subsamples to
roughly estimate the planet formation efficiency in the protoplanetary
disks that once existed around the stars in our sample.

\subsection{Logistic Regression}

We use logistic regression---a natural extension of linear regression for
probability---to obtain a first look at the relationship between host
star metallicity \& effective temperature and the probability of the
presence of a small planet candidate in the system $P_{\mathrm{host}}$.
We use the logistic regression model
\begin{eqnarray}\label{eqn:logistic_def}
P_{\mathrm{host}} & = & \frac{1}{1 + e^{-x}}, \\
x & = & \beta_0 + \beta_1 T_{\mathrm{eff}} + \beta_2 [\mathrm{M/H}],
\end{eqnarray}
and the \texttt{statsmodel.logit} \citep{genz04,statsmodel} implementation
of logistic regression.  We give the result of our calculation in
Table~\ref{tbl-3}.

\begin{deluxetable}{ccccc}
\tablecaption{Logistic Regression Results\label{tbl-3}}
\tablehead{
\colhead{Variable} & \colhead{Value} & \colhead{Uncertainty} & \colhead{$t$-statistic} & \colhead{$p$-value}}
\startdata
$T_{\mathrm{eff}}$ & $-7.6 \times 10^{-4}$ & $5.9 \times 10^{-4}$ & $-1.3$ & $0.20$ \\
$[\mathrm{M/H}]$  & $1.8$ & $0.65$ & $2.8$ & $0.010$
\enddata
\end{deluxetable}

We find that the coefficient for metallicity in the logistic regression
equation is positive and significantly different than zero, while
the coefficient for effective temperature is consistent with zero
(see Table~\ref{tbl-3}).  The implication is that planet occurrence
increases with host star metallicity and is insensitive to host star
effective temperature.  The coefficient of a continuous predictor
variable in a logistic regression model gives the expected change in
the natural logarithm of the odds ratio of the modeled outcome with a
one-unit change in that continuous predictor variable.  Since we will
subsequently find in the next subsection that the metallicity difference
between our metal-rich and metal-poor subsamples is about 0.3 dex, we use
our logistic regression model to estimate the effect of a 0.3 dex change
in $[\mathrm{M/H}]$ on the probability of finding a planet candidate in
a system $P_{\mathrm{host}}$.  When the probability of an event is small
and therefore $x$ must be small as well, the logistic regression function
is approximately an exponential regression $P=e^{x}$ and the coefficients
$\beta_{i}$ can be interpreted as the fractional change of the odds of an
event's occurrence.  We find that $P_{\mathrm{host}}([\mathrm{M/H}]+0.3)
= 1.69_{-0.30}^{+0.36}~P_{\mathrm{host}}([\mathrm{M/H}])$.  In words,
the probability that a late-type dwarf star was observed by Kepler
to host at least one small planet candidate increases by a factor of
about $1.69_{-0.30}^{+0.36}$ for a 0.3 dex change in $[\mathrm{M/H}]$
(a factor of two in $Z_{\star}$).

The logistic regression analysis handles single- and multiple-planet
systems in the same way and therefore does not account for multiplicity.
It does not control for the decrease in transit probability with semimajor
axis or the incompleteness of the Kepler DR25 planet candidate list.
It implicitly assumes that a star with no observed planet candidates is
equivalent to a star without planets.  This last assumption is only valid
in the parts of parameter space where planet occurrence is low (i.e.,
$P \lesssim 10$ days and $2~R_{\oplus} \lesssim R_{\mathrm{p}} \lesssim
5~R_{\oplus}$).  Because planet occurrence increases with both increasing
orbital period and with decreasing planet radius, it is important to
account for transit probability and Kepler DR25 completeness to explore
the connection between host star metallicity and small-planet occurrence
for the much more common long-period and/or small-radius planets.

\subsection{Occurrence as a Function of Metallicity, Orbital Period,
and Planet Radius}

To complement the logistic regression analysis in the previous subsection,
we calculate planet occurrence as a function of metallicity, orbital
period, and planet radius using the Kepler DR25 planet candidate list and
its completeness data products.  This approach takes into account planet
multiplicity and allows us to explore the connection between host star
metallicity and small-planet occurrence at longer periods and smaller
radii than the logistic regression approach.

We first separate our complete sample into metal-rich and metal-poor
subsamples by splitting at the metallicity that separates our planet
candidate-host sample into two nearly equal halves.  We split our complete
sample into two nearly equally sized subsamples to minimize the effects
of sample size differences.  We therefore set the dividing line at
$[\mathrm{M/H}] = -0.15$ as shown Figure~\ref{fig02}.  The resulting
metal-rich subsample has 74 planet candidates (49 planet candidate
hosts) and 1,299 non-planet-candidate hosts while the metal-poor
subsample has 76 planet candidates (50 planet-candidate hosts) and 2,096
non-planet-candidate hosts.  We find that the average metallicities
of our metal-rich and metal-poor samples are $[\mathrm{M/H}]= +0.0$
and $[\mathrm{M/H}]= -0.3$ respectively.  Because one star might host
multiple planet candidates, it is impossible to exactly divide the planet
candidate hosts such that the number of planet candidates in the two
subsamples are exactly equal.  If instead we split our complete sample
into two subsamples each with an equal number of stars, then the median
metallicity of the metal-rich subsample would decrease by 0.01 dex in
[M/H] and the median metallicity of the metal-rich subsample would remain
the same.

\begin{figure}
\plotone{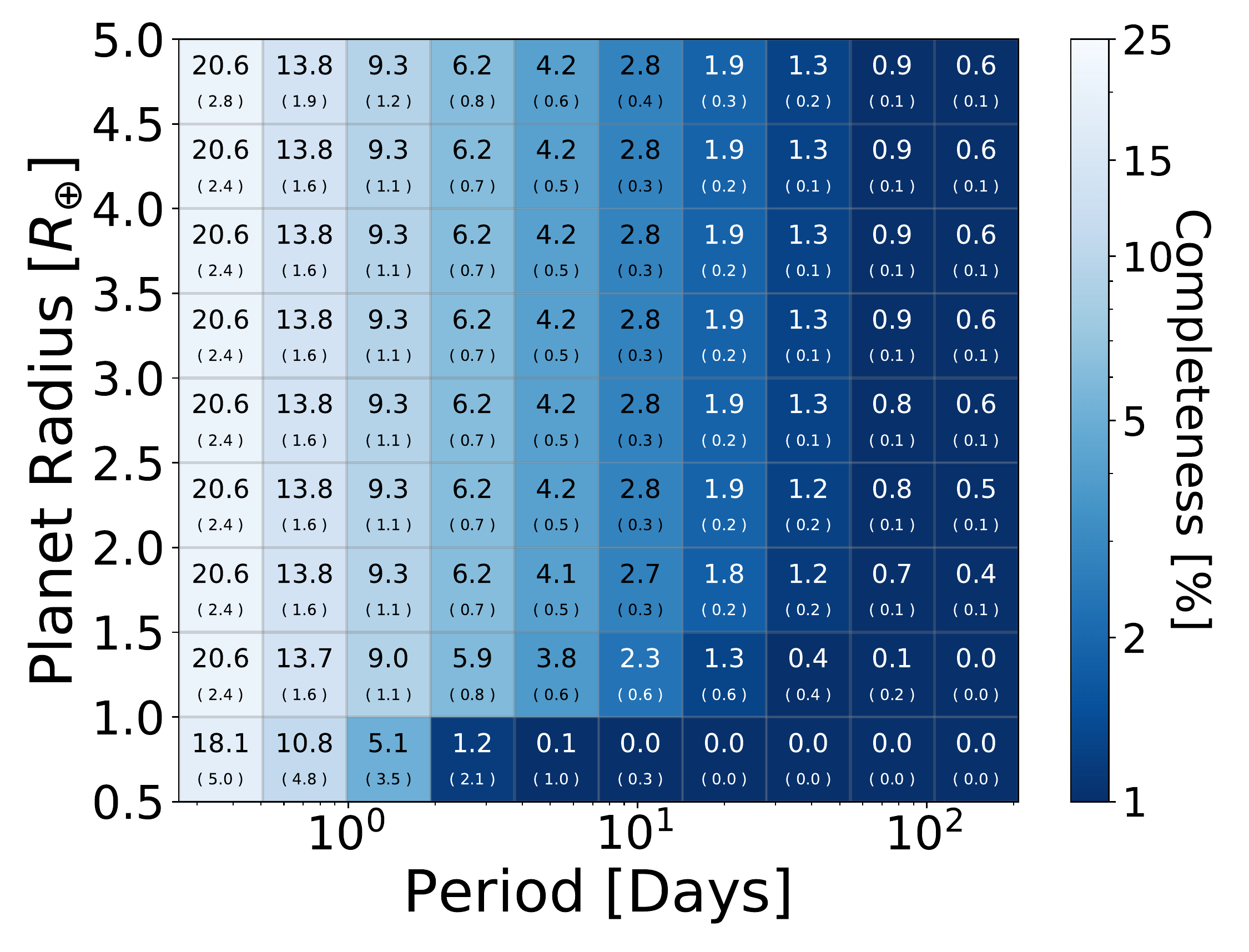}
\caption{Example completeness map for KIC 1577265. The values in every
cell denote the completeness and the scatter of completeness within the
cell in percent.\label{fig03}}
\end{figure}

To go from the observed frequency of planet candidates to their
underlying occurrence, it is necessary to divide the observed frequency
by its completeness.  For each star in both subsamples, we use the
\texttt{KeplerPORTS} software described in \citet{Burke+17} to estimate
the completeness of the Kepler Pipeline that produced the DR25 planet
candidate list as a function of orbital period and planet radius.
We present in Figure~\ref{fig03} an example completeness map for KIC
1577265 (a randomly selected star from our complete sample).

We combine individual completeness maps for all stars in each subsample
to obtain representative completeness maps for both the metal-rich and
metal-poor subsamples.  For each point in orbital period--planet radius
space, we take the average value of the completeness maps produced for
all stars in a given subsample.  We thereby obtain a representative
completeness map that corresponds to the typical completeness averaged
over an entire subsample.  Completeness maps that have smaller planet
radius cells than the typical planet candidate's radius uncertainty
are oversampled.  Since planet candidates only sparsely populate
orbital period--planet radius space and because planet candidate radius
uncertainties are non-negligible, we then resample the representative
completeness maps at lower resolution.  We evenly divide orbital
period--planet radius space in $\log_{10}{P}$ and $R_{\mathrm{p}}$.
For each cell, we take the representative completeness value to be
the median of all completeness estimates in that cell.  For example, in
Figure~\ref{fig03} the value in each cell is the median of all individual
completeness estimates in that cell from the initial higher-resolution
completeness map.

We compute the occurrence of planet candidates as a function of orbital
period and planet radius in both metal-rich and metal-poor subsamples
using their the representative completeness maps.  The occurrence in
each cell depends on the total number of observed planet candidates
$N_{\mathrm{PC}}$ and the total number of equivalent stars searched
$N_{\mathrm{\star}}$ in that cell.  We define $N_{\star}$ as the product
of our estimated representative completeness in that cell and the
total number of stars in a subsample.  Since there are uncertainties
in the measurement of each planet candidate's orbital period and
radius\footnote{Since the planet radius uncertainties provided in the
DR25 planet candidate list only include the effect of stellar radius
uncertainties, we calculated our own planet radius uncertainties
accounting for both transit depth and stellar radius uncertainties.},
we use a 1,000 iteration Monte Carlo simulation to distribute the impact
of an individual planet candidate detection across multiple cells using
a two-dimensional Gaussian kernel with a diagonal covariance matrix
with the 1-$\sigma$ period and radius uncertainties on the diagonal.
We then define $N_{\mathrm{PC}}$ as the number of counts in each cell
averaged over the Monte Carlo simulation.

We adopt a Bayesian framework to estimate planet occurrence $\eta$.
We model occurrence with a binomial likelihood and use a Beta distribution
prior $\mathrm{Beta}\left(\alpha,\beta\right)$.  In that situation, the
Beta distribution is a conjugate prior and the posterior distribution
of occurrence will be a Beta distribution that depends on the prior
parameters, $N_{\mathrm{PC}}$, and $N_{\star}$
\begin{eqnarray}\label{eqn:posterior}
P\left(\eta|N_{\mathrm{PC}},N_{\star}\right) & = & \\
&& \mathrm{Beta}\left(\alpha+N_{\mathrm{PC}},\beta+N_{\star}-N_{\mathrm{PC}}\right),\nonumber
\end{eqnarray}
where $\alpha$ and $\beta$ are parameters of the prior.  We assume a
weak uninformative prior with $\alpha = \beta = 1$.

\begin{figure*}
\plottwo{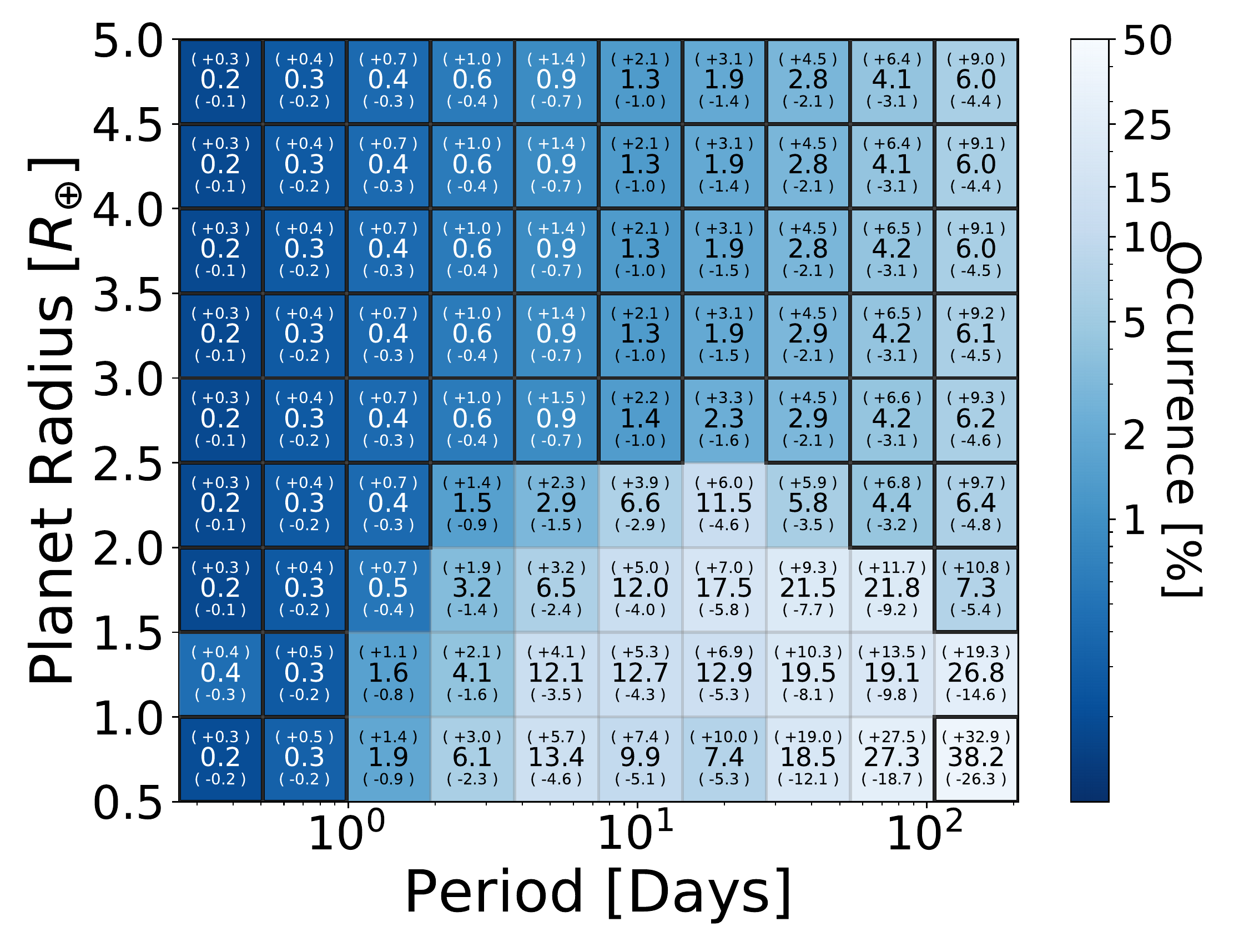}{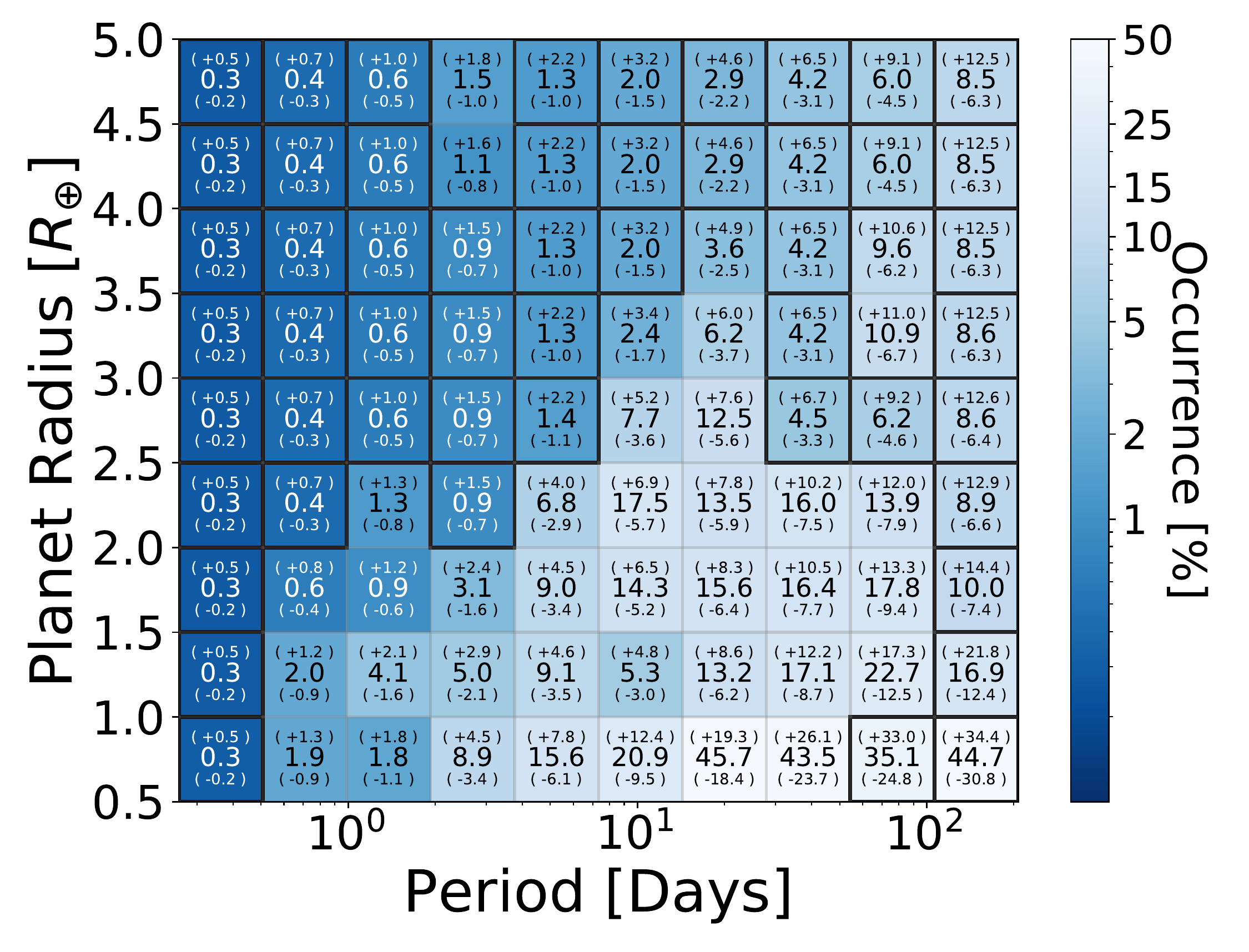}
\caption{Planet candidate occurrence in metallicity--period--planet
radius space with an uninformative prior.  The values in each cell are
the occurrence of planet candidates in that cell and its uncertainty.
All values are expressed as percents.  Cells with heavy borders have
no detected planet candidates. Left: small planet candidate occurrence
in our metal-poor subsample. Right: small planet candidate occurrence
in our metal-rich subsample.  Planet candidates are significantly more
common in the metal-rich subsample than in the metal-poor subsample.
The product of our samples' sizes and Kepler DR25's completeness
indicate that the amount of information implicit in a non-detection is
at least a factor of five larger than the signal weakly implied by our
prior for cells with $P \lesssim 100$ days and $R_{\mathrm{p}} \gtrsim
1~R_{\oplus}$.\label{fig04}}
\end{figure*}

\begin{figure}
\plotone{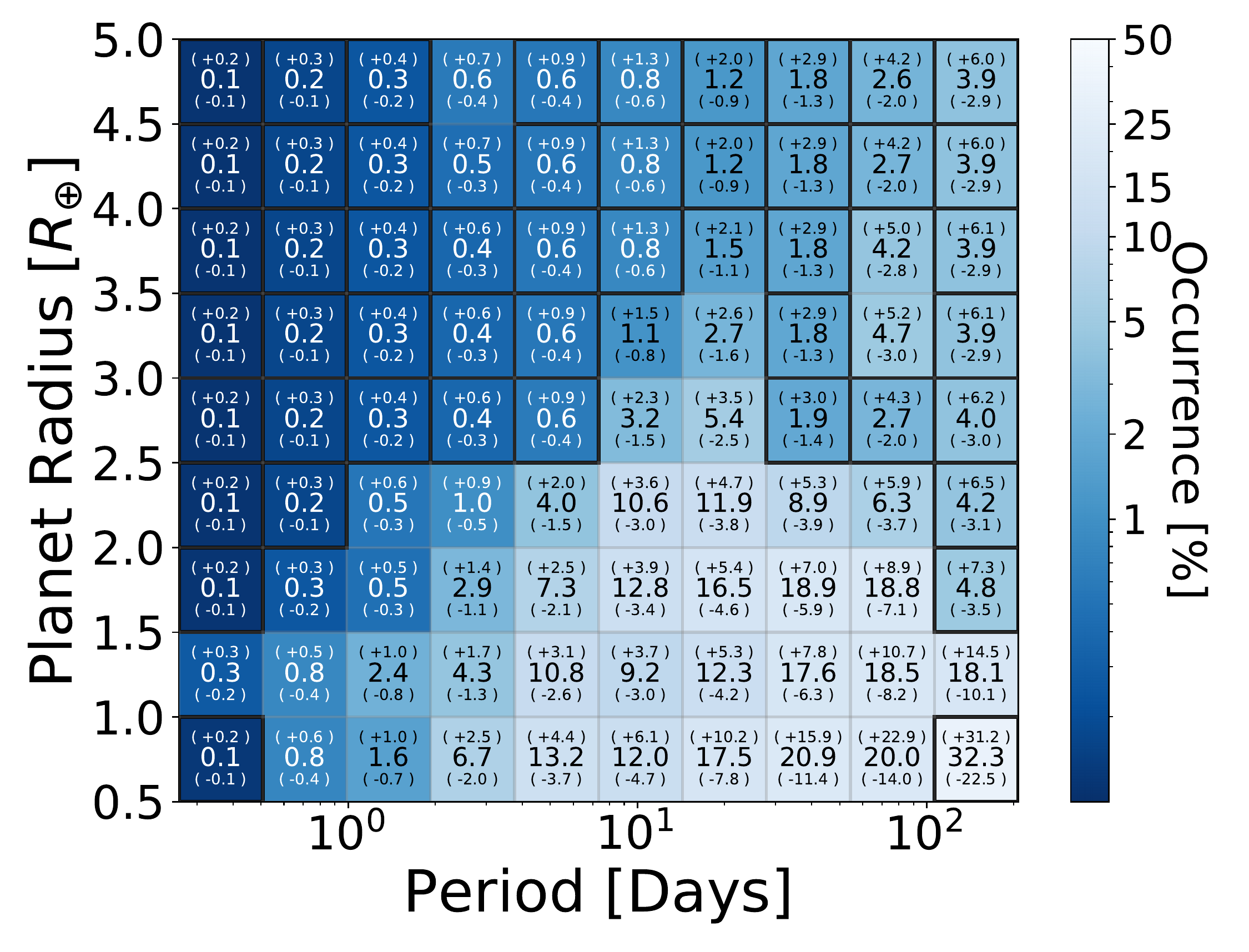}
\caption{Planet candidate occurrence as a function of period and
planet radius in our complete sample.  The values in each cell are
the occurrence of planet candidates in that cell and its uncertainty.
All values are expressed as percents. Cells with heavy borders have no
detected planet candidates.  The product of our samples' sizes and Kepler
DR25's completeness indicate that the amount of information implicit
in a non-detection is at least an order-of-magnitude larger than the
signal weakly implied by our prior for cells with $P \lesssim 100$
days and $R_{\mathrm{p}} \gtrsim 1~R_{\oplus}$.\label{fig05}}
\end{figure}

We plot the results of our occurrence calculations in Figures~\ref{fig04}
and \ref{fig05} and give them in tabular form in Table~\ref{tbl-4}.
Figure~\ref{fig04} shows planet candidate occurrence as a function of
orbital period and planet radius for both our metal-rich and metal-poor
subsamples, while Figure~\ref{fig05} shows planet candidate occurrence
as a function of orbital period and planet radius for our complete
sample. The differences in planet occurrence between the metal-rich
and metal-poor subsamples illustrate the effect of metallicity on small
planet formation: small planets are less common around metal-poor stars
than around metal-rich stars. We indicate cells with no detected planet
candidates in Table \ref{tbl-4} and with black borders in both Figures
\ref{fig04} and \ref{fig05}.  Our metal-rich and metal-poor subsamples
are large enough and Kepler DR25's completeness is high enough that
for cells with $P \lesssim 100$ days and $R_{\mathrm{p}} \gtrsim
1~R_{\oplus}$ the product of sample size and completeness is larger
than 10 (see Table~\ref{tbl-4}).  In this case, the signal implicit in
a non-detection is at least an a factor of five larger than the signal
weakly implied by our prior.

\begin{deluxetable*}{cccccccc}
\tablecaption{Occurrence of Small Planet Candidates in the Kepler Field
with Late-type Dwarf Primaries as a Function of Metallicity}
\tablehead{
\colhead{Planet Radius} & \colhead{Period} & \colhead{Occurrence } & \colhead{
$\mathrm{[M/H]~Description}$} &\colhead{PC Detection Flag} &\colhead{Completeness} &\colhead{Equivalent Number}\\
\colhead{} & \colhead{} & \colhead{} & \colhead{}  &\colhead{} &\colhead{} &\colhead{of Stars Searched}\\
\colhead{($R_{\oplus}$)} & \colhead{(days)} & \colhead{(\%)} & & & \colhead{(\%)} & }
\startdata
0.5-1.0 & 0.3-0.5 & ${0.2}^{+0.3}_{-0.2}$ & MP & 0 & 16.41 & 226\\
0.5-1.0 & 0.5-1.0 & ${0.3}^{+0.5}_{-0.2}$ & MP & 0 & 10.22 & 141\\
0.5-1.0 & 1.0-1.9 & ${1.9}^{+1.4}_{-0.9}$ & MP & 1 & 6.21 & 85\\
0.5-1.0 & 1.9-3.8 & ${6.1}^{+3.0}_{-2.3}$ & MP & 1 & 3.64 & 50\\
0.5-1.0 & 3.8-7.3 & ${13.4}^{+5.7}_{-4.6}$ & MP & 1 & 1.97 & 27\\
0.5-1.0 & 7.3-14.3 & ${9.9}^{+7.4}_{-5.1}$ & MP & 1 & 0.99 & 14\\
0.5-1.0 & 14.3-27.9 & ${7.4}^{+10.0}_{-5.3}$ & MP & 1 & 0.46 & 6\\
0.5-1.0 & 27.9-54.5 & ${18.5}^{+19.0}_{-12.1}$ & MP & 1 & 0.20 & 3\\
0.5-1.0 & 54.5-106.2 & ${27.3}^{+27.5}_{-18.7}$ & MP & 1 & 0.08 & 1\\
0.5-1.0 & 106.2-207.2 & ${38.2}^{+32.9}_{-26.3}$ & MP & 0 & 0.03 & 0\\
\enddata
\tablecomments{In the column ``[M/H] Description'' the strings ``MP'',
``MR'', and ``All'', correspond to our metal-poor, metal-rich, and
complete samples.  This table is published in its entirety in the
machine-readable format.\label{tbl-4}}
\end{deluxetable*}

\begin{figure*}
\plottwo{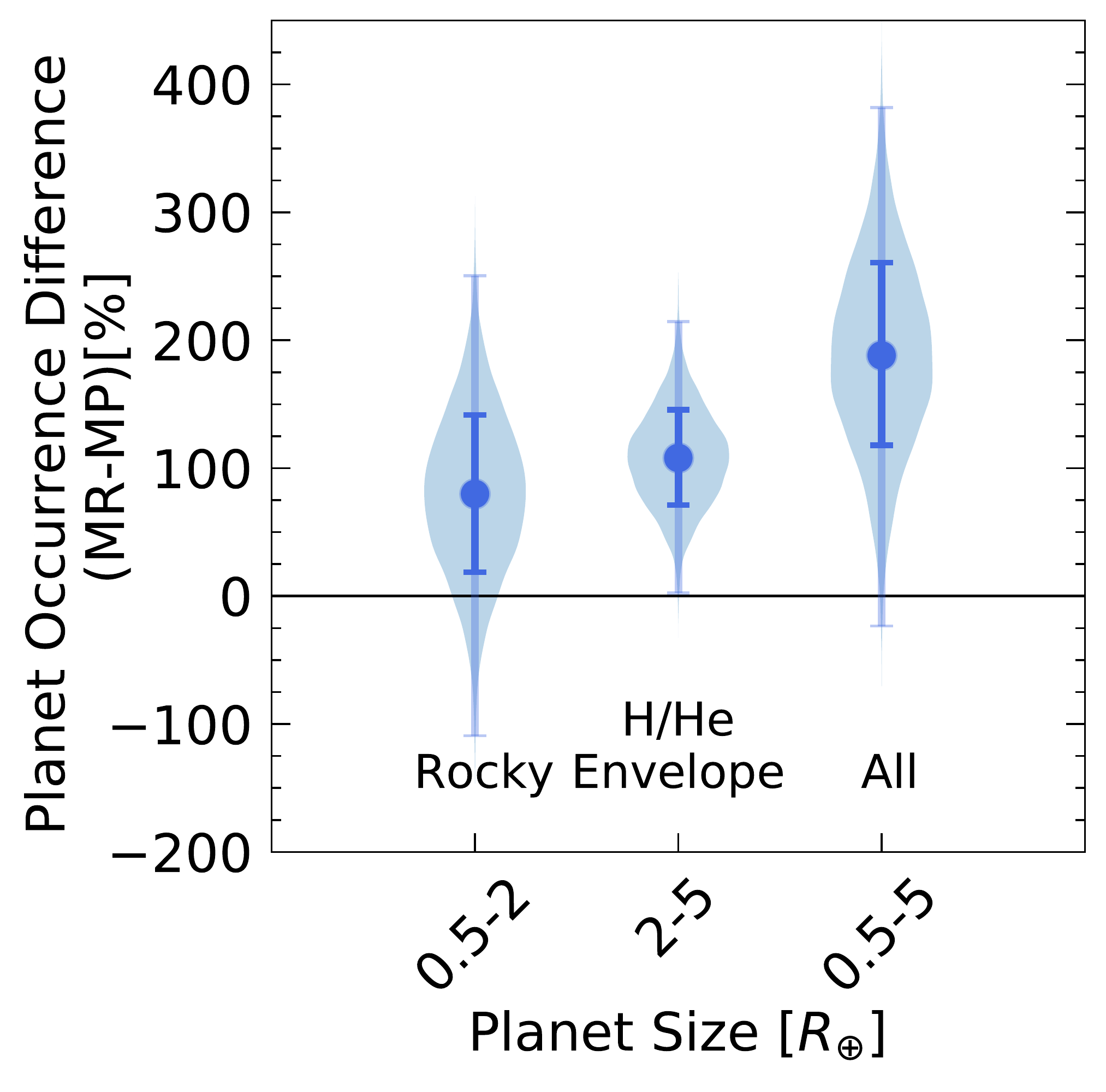}{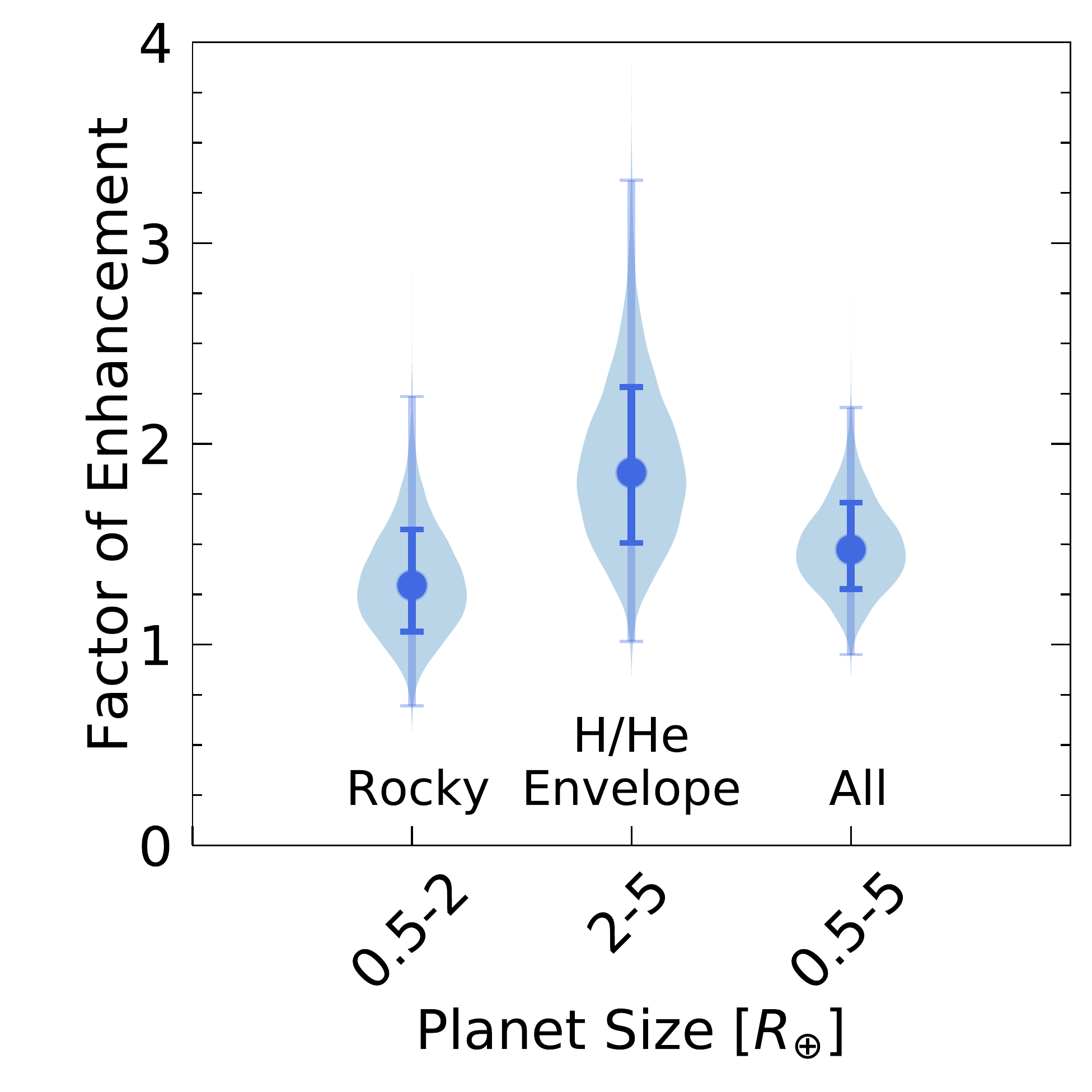}
\caption{Violin plots with differences in planet candidate occurrence
between metal-rich and metal-poor subsamples as a function of planet
radius.  The dark blue bars denote the 16th and 84th percentiles
(i.e., the 1-$\sigma$ region) while the light blue bars represents the
0.13th and 99.7th percentiles (i.e., the 3-$\sigma$ region).  Left:
planet occurrence difference as a function of planet radius.  Right:
factor of enhancement as a function of planet radius.  The occurrence of
planet candidates is significantly higher in our metal-rich subsample
both for the entire range of radii we study $0.5~R_{\oplus} \lesssim
R_{\mathrm{p}} \lesssim 5~R_{\oplus}$ and for H/He envelope planets
with $2~R_{\oplus} \lesssim R_{\mathrm{p}} \lesssim 5~R_{\oplus}$.
Our results are inconclusive for rocky planets with $0.5~R_{\oplus}
\lesssim R_{\mathrm{p}} \lesssim 2~R_{\oplus}$.\label{fig06}}
\end{figure*}

Planets with $R_{\mathrm{p}} \gtrsim 2~R_{\oplus}$
are almost certain to possess significant H/He envelopes
\citep[e.g.,][]{Rogers15,Chen+Kipping17}.  We therefore separately study
the difference in ``rocky'' ($0.5~R_{\oplus} \lesssim R_{\mathrm{p}}
\lesssim 2~R_{\oplus}$) and ``H/He envelope'' ($2~R_{\oplus} \lesssim
R_{\mathrm{p}} \lesssim 5~R_{\oplus}$) planet occurrence between our
metal-rich and metal-poor subsamples.  This $R_{\mathrm{p}} \approx
2~R_{\oplus}$  boundary also corresponds to the so-called ``Fulton
Gap'' \citep[e.g.,][]{Fulton+17,Fulton+18,Berger+2018}.  To faithfully
account for the effect of uncertainty on this calculation, we conduct
a Monte Carlo simulation.  For every cell of each of the metal-rich
and metal-poor subsamples, we sample planet candidate occurrence from
its posterior distribution.  For each cell, we then take the planet
candidate occurrence difference between the metal-rich and metal-poor
subsamples and sum the difference across all periods (excluding the
longest-period cells because of their sub-percent completeness levels).
We obtain a number that describes the cumulative differential occurrence
between metal-rich and metal-poor subsamples.  We repeat this process
10,000 times to fully sample the differential occurrence distribution.
We take the median and the 16th and 84th percentiles of the distribution
as the typical planet candidate occurrence difference and its associated
uncertainty.  We call this statistic our ``planet occurrence difference''
from here.  We visualize these results in Figure~\ref{fig06} and present
them in tabular form in Table~\ref{tbl-5}.

We calculate the enhanced occurrence of planets in the metal-rich
subsample relative to the occurrence of planets in the metal-poor
sample in one more way.  For every cell of each of the metal-rich and
metal-poor subsamples, we sample planet candidate occurrence from its
posterior distribution and sum over all cells in a subsample.  We divide
the summed occurrence calculated for the metal-rich subsample by the
summed occurrence calculated for the metal-poor subsample to calculate
a statistic we define as the occurrence ``factor of enhancement''.
We repeat this process 10,000 times to fully sample the factor of
enhancement distribution.  We report the median and the $16$th and $84$th
percentiles of the factor of enhancement distribution in Table~\ref{tbl-5}
and present it visually in Figure~\ref{fig06}.

\begin{deluxetable*}{lcc}
\tablecaption{Relative Planet Candidate Occurrence Statistics Observed
Between the Metal-rich and Metal-poor Subsamples}\label{tbl-5}
\tablehead{
\colhead{Category} & \colhead{Occurrence Difference} &
\colhead{Factor of Enhancement}\\
\colhead{} & \colhead{(\%)} &
\colhead{}}
\startdata
Rocky ($0.5~R_{\oplus} \lesssim R_{\mathrm{p}} \lesssim 2~R_{\oplus}$) & $80^{+62}_{-62}$ & $1.3^{+0.3}_{-0.2}$ \\
H/He Envelope ($2~R_{\oplus} \lesssim R_{\mathrm{p}} \lesssim 5~R_{\oplus}$) &  $108^{+38}_{-37}$& $1.9^{+0.4}_{-0.4}$ \\
All ($0.5~R_{\oplus} \lesssim R_{\mathrm{p}} \lesssim 5~R_{\oplus}$) &  $188^{+72}_{-70}$& $1.5^{+0.3}_{-0.2}$
\enddata
\end{deluxetable*}

For planet candidates over the complete range in planet radius we study
$0.5~R_{\oplus}\lesssim R_{\mathrm{p}} \lesssim 5~R_{\oplus}$, the
occurrence of planet candidates in the metal-rich subsample is a factor
of $1.5^{+0.3}_{-0.2}$ higher than in the metal-poor subsample.  For H/He
envelope planet candidates ($2~R_{\oplus} \lesssim R_{\mathrm{p}} \lesssim
5~R_{\oplus}$), the occurrence of planet candidates in the metal-rich
subsample is a factor of $1.9 \pm 0.4$ higher than in the metal-poor
subsample.  We note that the difference in $[\mathrm{M/H}]$ between our
metal-rich and metal-poor subsamples is about 0.3 dex or a factor of two
in $Z_{\star}$, so the occurrence of planets with $R_{\mathrm{p}} \gtrsim
2~R_{\oplus}$ grows roughly linearly with $Z_{\star}$ in our sample.
We also note that none of the planet candidates with $3~R_{\oplus}
\lesssim R_{\mathrm{p}} \lesssim 5~R_{\oplus}$ in our complete sample
were found in the metal-poor subsample.  We therefore conclude that for
late-type dwarf stars metallicity is an important parameter in planet
occurrence calculations that should not be neglected for planets with
$R_{\mathrm{p}} \gtrsim 2~R_{\oplus}$.  Studies of small planet occurrence
around late-type dwarfs using K2 or TESS data should therefore be sure
to control for metallicity.

For rocky planet candidates with $0.5~R_{\oplus} \lesssim R_{\mathrm{p}}
\lesssim 2~R_{\oplus}$, the occurrence of planet candidates in the
metal-rich subsample is a factor of $1.3^{+0.3}_{-0.2}$ higher than in
the metal-poor subsample.  Given the 1-$\sigma$ significance of this
observation, we cannot confirm or reject a relationship between host
star metallicity and planet candidate occurrence.

To compare with previous estimates of the occurrence of small planets
around late-type dwarfs in the Kepler field, we calculated the occurrence
of planet candidates in the ranges $0.5~R_{\oplus} \lesssim R_{\mathrm{p}}
\lesssim 4~R_{\oplus}$ and $1~R_{\oplus} \lesssim R_{\mathrm{p}} \lesssim
4~R_{\oplus}$ with orbital period $P < 207$ days in our complete sample.
We find that in these radius ranges a late-type dwarf in our complete
sample hosts $3 \pm 0.3$ and $4.4_{-0.4}^{+0.5}$ planets respectively.
These results are consistent with those reported in other studies of
small-planet occurrence in the late-type dwarf stellar population in the
Kepler field \citep[e.g.,][]{Dressing+2015,Hsu+20}.  This consistency
supports the accuracy of our occurrence calculations.

\subsection{Formation Efficiency of Small Planets}\label{section:efficiency}

The fraction of planet-making material present in a protoplanetary disk
during the epoch of planet formation that ends up sequestered in planets
can be thought of as that disk's planet formation efficiency.  As we
described in Section 1, pebble accretion is expected to be inefficient
with planet formation efficiencies below 10\%.  On the other hand,
the apparent planet formation efficiency in the solar system was much
higher.  We therefore estimate the planet formation efficiency in the
Kepler field's late-type dwarf stellar population in an attempt to
observationally constrain the planet formation process.

To estimate planet formation efficiency, we need both the expectation
value for the mass in planets today as well as the total amount of
planet-making material that was available in the young disk.  To calculate
the former, we use the small-planet occurrence we estimated above
combined with the mass--radius relation presented in \citet{Ning+2018} and
implemented in the \texttt{MRExo} package \citep{Kanodia+2019}.  We note
that the \citet{Ning+2018} mass--radius relation does not distinguish
between a planet's mass in metals and its mass in hydrogen and helium.
Since the masses of planets smaller than Neptune are dominated by
their metal mass, this should only bias our results by about 10\%
\citep[e.g.,][Schlaufman \& Halpern 2020 submitted]{Podolak+19}.  We use
a Monte Carlo simulation in which we sample the occurrence in each cell
of the maps presented in Figures~\ref{fig04} and \ref{fig05} from the
occurrence posterior in each cell.  We next multiply that occurrence by
the mass predicted by the \citet{Ning+2018} mass--radius relation at the
radius of the cell's midpoint.  We then sum the product of occurrence
and mass for each cell over an entire occurrence map (excluding the
longest-period cells because of their sub-percent completeness levels).
We save the resulting estimate of the expectation value for the total
mass in planets and repeat the process 10,000 times.  We perform a
similar simulation for the complete sample as well as the metal-rich
and metal-poor subsamples.  We report the expected mass in planets for
all three samples in Table~\ref{tbl-6}.

\begin{deluxetable}{lc}
\tablecaption{Expected Mass in Planets as a Function of
Metallicity}\label{tbl-6}
\tablehead{
\colhead{Sample} & \colhead{Expected Mass in Planets}\\
\colhead{} & \colhead{($M_{\oplus}$)} }
\startdata
Metal-poor  &  $16.5_{-1.8}^{+0.6}$ \\
Metal-rich  &  $24.5_{-2.5}^{+0.9}$ \\
Complete    &  $13.9_{-1.2}^{+0.5}$
\enddata
\end{deluxetable}

To calculate the total amount of planet-making material that was
available in the protoplanetary disks once present around the late-type
dwarfs in the Kepler field, we use the same back-of-the-envelope
calculation described in Section 1.  We assume $Z_{\mathrm{disk}}
= Z_{\star} = (0.0070, 0.010, 0.014)$ for the metal-poor, complete,
and metal-rich samples.  We use the \citet{Andrews2013} relation with
$M_{\mathrm{disk}}/M_{\star} = 0.01$ for an early M dwarf with $M_{\star}
= 0.6~M_{\odot}$.  We therefore estimate the amount of planet-making
material available in the protoplanetary disks around the stars in
our metal-poor, complete, and metal-rich samples as $14~M_{\oplus}$,
$20~M_{\oplus}$, and $28~M_{\oplus}$.

We find planet formation efficiencies in excess of 50\%.  The implication
is that either planet formation is very efficient or that the small
planet candidates observed around the Kepler field's late-type dwarf
stellar population formed in disks more massive than the average disks
observed by \citet{Andrews2013}.  This could be because their parent
protoplanetary disks were preferentially drawn from the high-mass
side of the \citet{Andrews2013} distribution or because these planet
candidates formed in younger and therefore more massive disks than those
observed by \citet{Andrews2013}.  We also ignore the uncertainty in the
\citet{Ning+2018} mass--radius relation.  Nevertheless, our observation's
preference for massive disks is similar to that suggested in the
minimum-mass extrasolar nebula scenario proposed by \citet{Chiang+2013}
and expanded by \citet{Dai+2020}.  It is important to note that our
estimated planet formation efficiency is limited to planets falling
within our occurrence maps, or $R_{\mathrm{p}} \lesssim 5~R_{\oplus}$
and $P \lesssim 200$ days.

\section{Discussion}\label{section:discussion}

The occurrence of small planet candidates in the union of our
metal-poor and metal-rich subsamples is consistent with the results
of previous studies.  \citet{Dressing+2015} found that M dwarfs
host on average $2.5\pm 0.2$ planets with planet radii $1~R_{\oplus}
\lesssim R_{\mathrm{p}} \lesssim 4~R_{\oplus}$ and orbital period $P <
200$ days.  Our planet candidate occurrence of $3.0\pm 0.3$ planets per
late-type dwarf for the same period and radius range is consistent with
the \citet{Dressing+2015} estimate.  More recently, \citet{Hsu+20}
used a Bayesian framework to calculate M dwarf planet occurrence
for planet candidates with $0.5~R_{\oplus} \lesssim R_{\mathrm{p}}
\lesssim 4~R_{\oplus}$ and orbital period $0.5~\mathrm{days} < P <
256$ days.  Their planet occurrence ranges from $4.8^{+0.7}_{-0.6}$ to
$8.9^{+1.2}_{-0.9}$ planets per M dwarf depending on the choice of prior.
Our estimated occurrence is $4.4^{+0.5}_{-0.4}$ for planets with radii
$0.5~R_{\oplus} \lesssim R_{\mathrm{p}} \lesssim 4~R_{\oplus}$ and orbital
period $0.5~\mathrm{days} < P < 207$ days.  Our results are consistent
with those of \citet{Hsu+20}, though we use a slightly smaller maximum
period due to the large and uncertain completeness corrections required
for $P \gtrsim 200$ days.

We find significant increases in planet occurrence with metallicity
over both the entire range of planet radii we study ($0.5~R_{\oplus}
\lesssim R_{\mathrm{p}} \lesssim 5~R_{\oplus}$) and over the range of
radii indicative of planets with significant H/He envelopes ($2~R_{\oplus}
\lesssim R_{\mathrm{p}} \lesssim 5~R_{\oplus}$).  We find period-averaged
occurrences in the metal-rich samples higher than the occurrences
observed in the metal-poor samples by a factor $1.5_{-0.2}^{+0.3}$
for $0.5~R_{\oplus} \lesssim R_{\mathrm{p}} \lesssim 5~R_{\oplus}$ and a
factor of $1.9 \pm 0.4$ for $2~R_{\oplus} \lesssim R_{\mathrm{p}} \lesssim
5~R_{\oplus}$.  These factor-of-two enhancements are significant at more
than the 2-$\sigma$ level.  Since the average photometric metallicities
of the metal-rich and metal-poor subsamples differ by about 0.3 dex in
$[\mathrm{M/H}]$ (or a factor of two in $Z_{\star}$), the occurrence of
small planets overall and gas-rich planets specifically scales linearly
with metallicity.  This linear scaling applies at least in the thin
disk metallicity range probed by Kepler during its prime mission ($-0.5
\lesssim [\mathrm{M/H}] \lesssim +0.5$).  The amplitude and significance
of this effect implies that future studies of small planet occurrence
around thin disk late-type dwarf stars using K2 or TESS data should
control for the effect of metallicity on occurrence estimates.

Our results are inconclusive for rocky planets with $0.5~R_{\oplus}
\lesssim R_{\mathrm{p}} \lesssim 2~R_{\oplus}$.  We find a period-averaged
occurrence in the metal-rich sample higher than the occurrence observed
in the metal-poor sample by a factor $1.3_{-0.2}^{+0.3}$.  This hint
of an enhancement is only significant at the 1-$\sigma$ level.  We are
therefore unable to confirm a relationship between metallicity and planet
occurrence for rocky planets.

The lack of a statistically significant relationship between metallicity
and occurrence in the range $0.5~R_{\oplus} \lesssim R_{\mathrm{p}}
\lesssim 2~R_{\oplus}$ could be due to Kepler's low completeness for
small planets.  It could also be the case that there is no relationship
between metallicity and planet occurrence for the smallest planets.
We assert that the former is the best explanation.  Since the relationship
between planet occurrence and metallicity is set during the era of
planet formation, the subsequent atmospheric evolution of a planetary
system cannot alter the relation.  If planets with $0.5~R_{\oplus}
\lesssim R_{\mathrm{p}} \lesssim 2~R_{\oplus}$ are the leftover cores
of larger planets that were stripped of their H/He envelopes, then the
dependence of occurrence on metallicity should be the same for both
rocky and gas-rich planets.  In other words, the lack of a relationship
between metallicity and occurrence for the smallest planets that cannot
be attributed to low completeness would require that the small planets
observed by Kepler around late-type dwarfs formed like terrestrial
planets without significant H/He envelopes.  We argue that a more precise
quantification of the relationship between metallicity and small planet
occurrence should be a priority for K2 and TESS planet occurrence studies.

We confirm the reality of the relation between metallicity and
small-planet occurrence for late-type dwarf stars first noted by
\citet{Schlaufman+Laughlin2010,Schlaufman+Laughlin2011}.  Our use of
a vetted photometric metallicity relation and removal of all giant
stars from our non-planet-candidate-host sample using both Kepler
asteroseismology and Gaia DR2 parallaxes answers the criticisms of
the \citet{Schlaufman+Laughlin2011} result made by \citet{Mann+2012}.
In accord with \citet{Schlaufman+Laughlin2010}, we find that the average
metallicity of late-type dwarfs in the Kepler field is $[\mathrm{M/H}]
\approx -0.15$.

Before Kepler, the relationship between small-planet occurrence
and solar-type host star metallicity had only been explored with
Doppler-discovered planets.  Those studies suggested that the connection
between host star metallicity and small-planet occurrence was much
weaker than for giant planets \citep[e.g.,][]{Sousa2008,Mayor2011}.
Subsequent analyses of large samples of transit-discovered small
planets have produced mixed results.  A majority of those analyses
did not find significant metallicity offsets between stars with
and without transiting planet candidates, at least in the range of
metallicity probed by Kepler $-0.5 \lesssim [\mathrm{M/H}] \lesssim
+0.5$ \citep[e.g.,][]{Schlaufman+Laughlin2011,Buchhave+2012,
Buchhave+2015,Schlaufman2015}.  On the other hand, some studies
have found evidence supporting such a connection for medium-sized
planets \citep[e.g.,][]{Buchhave+2014,Wang+Fischer2015,Courcol+2016}.
This latter dependence has been reaffirmed by analyses making use of
large samples of spectroscopic stellar parameters for Kepler-field
stars based on low-resolution optical spectra from the Large Sky Area
Multi-Object Fibre Spectroscopic Telescope (LAMOST) and its massive sky
survey \citep[e.g.,][]{Zhu+2016,Dong+2018}.

None of the studies listed in the paragraph above have taken into
account Kepler's completeness and therefore could not fully explore the
connection between host star metallicity and small planet occurrence.
\citet{Petigura+18} was the first to account for completeness and
found for solar-type host stars that the occurrence of planets
with $1.7~R_{\oplus} \lesssim R_{\mathrm{p}} \lesssim 4~R_{\oplus}$
doubles as stellar metallicity increases over the range  $-0.4 \lesssim
[\mathrm{M/H}] \lesssim +0.4$ (or a factor of six in $Z_{\star}$).
We find a factor of two change the occurrence of $2~R_{\oplus} \lesssim
R_{\mathrm{p}} \lesssim 5~R_{\oplus}$ planets over a smaller range of
metallicity $-0.3 \lesssim [\mathrm{M/H}] \lesssim +0.0$ (or a factor
of two in $Z_{\star}$).  Our study therefore verifies the theoretical
expectation that the connection between host star metallicity and
small-planet occurrence should be stronger for late-type dwarfs than
for solar-type stars.

With the connection between small-planet occurrence and late-type
dwarf host star metallicity now firmly established, it is possible
to predict the occurrence and properties of small planets around
late-type dwarf stars as a function of stellar metallicity and mass.
Assuming a planet-formation efficiency of 50\% and that the amount
of planet-making material available in a protoplanetary disk with
$M_{\mathrm{disk}}/M_{\star} = 0.01$ scales with stellar mass and
metallicity, during the epoch of planet formation there will be less
than $9~M_{\oplus}$ of planet-making material in the disk around a
$M_{\star} \approx 0.6~M_{\odot}$ early-type M dwarf with $[\mathrm{M/H}]
\lesssim -0.5$.  This meager amount of planet-making material is barely
sufficient to make even an Earth-composition $1.7~R_{\oplus}$ planet
\citep[e.g.,][]{Zeng+2019}.  Using the same assumptions for a late-type
M dwarf like 2MASS J23062928-0502285 (TRAPPIST-1) with $M_{\star}
\approx 0.08~M_{\odot}$ and $[\mathrm{M/H}] \approx 0$,  there will be
about $4~M_{\oplus}$ of planet-making material available.  Assuming an
Earth-like composition for the seven known TRAPPIST-1 planets implies
a total mass of about $7~M_{\oplus}$ \citep{Gillon+2016,Gillon+2017}.
We therefore predict that TRAPPIST-1 is metal-rich and/or that its
planetary system formed early in a massive protoplanetary disk.  In either
case, TRAPPIST-1 like systems should be very uncommon in future planet
occurrence studies of late-type M dwarfs like \citet{Sestovic+20}.

For early M dwarfs in the Kepler field, we estimate that more than 50\% of
the planet-making material initially present in their protoplanetary disks
was sequestered in planets.  Even if we assume disks an order of magnitude
more massive than the typical disk observed by \citet{Andrews2013}, this
is still larger than the expected $\lesssim\!\!10$\% of planet-making
material locked up in planets as a result of pebble accretion.  While both
our inability to differentiate between solid and gas masses for the small
planets in our sample and our exclusion of giant planets may bias our
planet formation efficiencies, we argue that these effects are small.
Neptune-size or smaller planets have less than 10\% of their mass in H/He
envelope, while giant planets occur around only a few percent of early
M dwarfs \citep[e.g.,][]{Podolak+19,Johnson+2010}.  We therefore argue
that the high planet formation efficiencies observed by \citet{Dai+2020}
and ourselves hint at planetesimal accretion as the main formation
channel for the small planets around early M dwarfs in the Kepler field.
While our planet formation efficiency calculation has large uncertainties
and may be systematically biased, we hope that future analyses of
the occurrence of small planets around low-mass stars may be able to
improve the estimation of planet formation efficiencies and thereby more
confidently differentiate between pebble and planetesimal accretion.

Even though our logistic regression analysis cannot account for the
important issues of completeness and multiple-planet systems, it still
provides an estimate of the strength of the small-planet occurrence--host
star metallicity relation that is consistent with our more careful
occurrence analysis.  Specifically, our logistic regression analysis
indicates that a 0.3 dex increase in $[\mathrm{M/H}]$ increases planet
occurrence in the range $0.5~R_{\oplus} \lesssim R_{\mathrm{p}} \lesssim
5~R_{\oplus}$ by a factor of $1.7 \pm 0.3$.  The more robust occurrence
calculation accounting for completeness and multiple planet systems
suggests a factor of $1.5_{-0.2}^{+0.3}$ increase for the same change
in metallicity.  These two estimates are consistent at the 1-$\sigma$
level.  The reason for this agreement is that the assumptions of the
logistic regression analysis are reasonable in regions of parameter space
where planet occurrence is low.  In other words, a logistic regression
analysis is an easy way to explore the dependence of planet occurrence
on other system parameters at short orbital periods and/or at relatively
large planet masses or sizes where planets are intrinsically uncommon
(see Figure~\ref{fig04}).  As most planets discovered by K2 and TESS are
on short-period orbits because of their limited observation durations,
we suggest that a logistic regression analysis could easily be used
to explore the dependence of planet occurrence on metallicity or other
system parameters among K2 or TESS discoveries even without accounting
for completeness.

\section{Conclusion}\label{section:conclusion}

We find that the occurrence of small planets around early M dwarfs in the
Kepler field increases linearly with host star metallicity $Z_{\star}$
for planets with H/He envelopes in the radius range $2~R_{\oplus} \lesssim
R_{\mathrm{p}} \lesssim 5~R_{\oplus}$ and $-0.3 \lesssim [\mathrm{M/H}]
\lesssim +0.0$.  We are unable to confirm or reject a relationship
between planet occurrence and host star metallicity for rocky planets
with $0.5~R_{\oplus} \lesssim R_{\mathrm{p}} \lesssim 2~R_{\oplus}$.
Similar analyses have shown an analogous but weaker increase in planet
occurrence with metallicity for solar-type stars in a similar range
of host star metallicity and period.  These observations confirm the
theoretical expectation that the small-planet occurrence--host star
metallicity relation should be stronger for low-mass stars.  Our results
provide a hint that planetesimal accretion should be preferred to pebble
accretion as the driving process for the formation of $2~R_{\oplus}
\lesssim R_{\mathrm{p}} \lesssim 5~R_{\oplus}$ planets around early M
dwarfs in the Kepler field.  We predict that even rocky planets with
$R_{\mathrm{p}} \gtrsim 1.7~R_{\oplus}$ or $R_{\mathrm{p}} \gtrsim
1.5~R_{\oplus}$ should be rare around early M dwarfs with $[\mathrm{M/H}]
\lesssim -0.5$ or late M dwarfs with $[\mathrm{M/H}] \lesssim +0.0$.
We argue that future small planet occurrence calculations for M dwarfs
targeted by K2 and/or TESS should control for metallicity.

\acknowledgements
We thank Hsiang-Chih Hwang, Bin Ren, Josh Winn, and Winston Wu
for useful discussions.  This paper includes data collected by the
Kepler mission.  Funding for the Kepler mission is provided by the
NASA Science Mission directorate.  This research has made use of the
NASA Exoplanet Archive, which is operated by the California Institute
of Technology, under contract with the National Aeronautics and Space
Administration under the Exoplanet Exploration Program.  Some/all of the
data presented in this paper were obtained from the Mikulski Archive
for Space Telescopes (MAST).  STScI is operated by the Association
of Universities for Research in Astronomy, Inc., under NASA contract
NAS5-26555.  This research has made use of the NASA/IPAC Infrared
Science Archive, which is funded by the National Aeronautics and Space
Administration and operated by the California Institute of Technology.
This publication makes use of data products from the Wide-field Infrared
Survey Explorer, which is a joint project of the University of California,
Los Angeles, and the Jet Propulsion Laboratory/California Institute
of Technology, and NEOWISE, which is a project of the Jet Propulsion
Laboratory/California Institute of Technology. WISE and NEOWISE
are funded by the National Aeronautics and Space Administration.
This work has made use of data from the European Space Agency (ESA)
mission {\it Gaia} (\url{https://www.cosmos.esa.int/gaia}), processed
by the {\it Gaia} Data Processing and Analysis Consortium (DPAC,
\url{https://www.cosmos.esa.int/web/gaia/dpac/consortium}).  Funding for
the DPAC has been provided by national institutions, in particular the
institutions participating in the {\it Gaia} Multilateral Agreement.
This research has made use of NASA's Astrophysics Data System.  This
research has made use of the SIMBAD database, operated at CDS, Strasbourg,
France \citep{Wenger+2000}.  This research has made use of the VizieR
catalogue access tool, CDS, Strasbourg, France (DOI: 10.26093/cds/vizier).
The original description of the VizieR service was published in 2000,
A\&AS 143, 23 \citep{Ochsenbein+2000}.  This project was developed in
part at the 2018 NYC Gaia Sprint, hosted by the Center for Computational
Astrophysics of the Flatiron Institute in New York City, New York.

\vspace{5mm}
\facilities{Exoplanet Archive, Gaia, Kepler, IRSA, MAST, NEOWISE, WISE}

\software{\texttt{Astropy} \citep{Astropy13,PriceWhelan18},
          \texttt{KeplerPORTS} \citep{Burke+17},
          \texttt{MRExo} \citep{Kanodia+2019},
          \texttt{pandas} \citep{pandas},
          \texttt{R} \citep{R20},
          \texttt{statsmodel} \citep{genz04, statsmodel}
          }

\bibliography{ms}
\bibliographystyle{aasjournal}
\end{CJK*}
\end{document}